\documentclass{epl}

\usepackage[T1]{fontenc}
\usepackage{epsfig}

\title{Mott insulator to superfluid transition of ultracold bosons in an
       optical lattice near a Feshbach resonance}   
\shorttitle{Mott-superfluid transition of ultracold bosons}
\author{K. Sengupta\inst{1} and N. Dupuis\inst{2}}
\institute{ \inst{1} Department of Physics, Yale university, New Haven,
CT-06520-8120 \\  
\inst{2} Laboratoire de Physique des Solides, CNRS UMR 8502, 
  Universit\'e Paris-Sud, 91405 Orsay, France }
\date{December 9, 2004}


\begin{document}


\maketitle

\begin{abstract}  
We study the phase diagram of  ultracold bosons in an optical
lattice near a Feshbach resonance. Depending on the boson density,
the strength of the optical lattice potential and the detuning
from resonance, the ground state can be a Mott insulator, a
superfluid phase with both an atomic and a molecular condensate,
or a superfluid phase with only a molecular condensate. Mott
insulator to superfluid transitions can be induced either by
decreasing the strength of the optical lattice potential or by
varying the detuning from the Feshbach resonance. Quite generally,
we find that for a commensurate density the ground-state may undergo 
several insulator-superfluid or superfluid-insulator transitions as the
magnetic field is varied through the resonance. 
\end{abstract}


Recent experiments on ultracold trapped atomic gases have opened
a new window onto the phases of quantum matter \cite{Greiner02}.
A gas of bosonic atoms in an optical or magnetic trap has been
reversibly tuned  between superfluid (SF)
and insulating ground states by varying the strength of a
periodic potential produced by standing optical waves.
This transition has been explained on the basis of the Bose-Hubbard
model with on-site repulsive interactions and hopping between
nearest neighboring sites of the lattice
\cite{Fisher89}.  As long as the atom-atom interactions are
small compared to the hopping amplitude, the ground state remains
superfluid. In the opposite limit of a strong lattice potential, the
interaction energy dominates and the ground state is a Mott insulator
(MI) when the density is commensurate, with an integer number of atoms
localized at each lattice site. 
Another aspect of trapped bosonic
atoms that has been studied extensively in recent years (without
the optical lattice) is the dramatic increase in the effective atom-atom
interaction due to the presence of a Feshbach resonance
\cite{Inouye98,Duine03}. Near the resonance, which can be
experimentally realized by tuning an external magnetic field, the
atoms can resonantly collide to form molecular bound states.
Remarkable phenomena associated with a Feshbach
resonance, including coherent atom-molecule oscillations and
collapse of the Bose-Einstein condensate, have been experimentally
observed \cite{Duine03}.

In this Letter, we study a Bose gas in the presence of an
optical lattice near a Feshbach resonance. Starting from an
effective Bose-Hubbard Hamiltonian, we first
obtain the various Mott insulating states that are stable in the limit
of a strong optical lattice. We then study the stability of these
states by deriving an effective Landau-Ginzburg theory for the
MI-SF transition.  Depending on the boson density, the
strength of the optical lattice 
potential and the detuning from resonance, the ground state can be a
MI, a SF phase with both an atomic and a 
molecular condensate, or a SF phase with only a molecular
condensate. MI-SF transitions can be induced 
either by decreasing the strength of the optical lattice potential or
by varying the detuning from the Feshbach resonance. For odd
commensurate densities, Mott phases are found to be always unstable to
superfluidity on the molecular side of the Feshbach resonance. For
even commensurate densities, Mott phases can be stable on both sides
of the resonance and may undergo several MI-SF or SF-MI
transitions as the magnetic field is varied through the resonance. 

Bosons in an optical lattice near a Feshbach resonance can be
described by the generalized Bose-Hubbard Hamiltonian
\begin{eqnarray}
H &=& \sum_{\sigma=a,m} \Bigl[ 
-t_\sigma \sum_{\langle {\bf r},{\bf r}'\rangle} (\psi^\dagger_\sigma({\bf r})
\psi_\sigma({\bf r}') + {\rm h.c.}) 
-\mu_\sigma \sum_{\bf r}
\psi^\dagger_\sigma({\bf r}) \psi_\sigma({\bf r})
\nonumber \\ && 
+ \frac{U_\sigma}{2} \sum_{\bf r} \psi_\sigma^\dagger({\bf
  r})\psi_\sigma^\dagger({\bf r}) \psi_\sigma({\bf r})
\psi_\sigma({\bf r}) \Bigr] \nonumber \\
&& + U_{am} \sum_{\bf r} \psi_a^\dagger({\bf
  r})\psi_m^\dagger({\bf r}) \psi_m({\bf r}) \psi_a({\bf r}) 
- \alpha \sum_{\bf r} (\psi_m^\dagger({\bf r}) \psi_a({\bf r})\psi_a({\bf r}) +
{\rm h.c.}) .
\label{ham1}
\end{eqnarray}
$\psi_\sigma,\psi^\dagger_\sigma$ are bosonic operators for atoms
($\sigma=a$) and molecules ($\sigma=m$). The discrete variable ${\bf
  r}$ labels the different sites (i.e. minima) of the optical
lattice. $\langle {\bf r},{\bf r}'
\rangle$ denotes a sum over nearest sites. The optical lattice is
assumed to be bipartite with 
coordination number $z$. $t_\sigma$ is the
intersite hopping amplitude, and $U_\sigma$ the amplitude of the
atom-atom ($\sigma=a$) or molecule-molecule ($\sigma=m$) interaction. 
$U_{am}$ gives the amplitude of atom-molecule
interactions. $\mu_a=\mu$ and $\mu_m=2\mu-\nu$ are the chemical
potential for atoms and molecules. The chemical potential $\mu$ fixes
the total number of atoms (free or bound into molecules). $\nu$ is
related to the energy of a molecular bound state and is a function of
the applied magnetic field. A negative detuning from resonance
($\nu<0$) favors the formation of molecules. The last term in Eq.~(\ref{ham1})
describes the atom-molecule conversion and the parameter $\alpha$
depends on the resonance width \cite{Inouye98,Duine03}.
The Hamiltonian (\ref{ham1}) can be derived from a microscopic
model  starting from the bare
atom-molecule Hamiltonian in the presence of the optical lattice
\cite{Duine03} and expanding the operators $\psi_a,\psi_m$ on the
lowest-band Wannier states \cite{note1}. We
shall not attempt to calculate the explicit values of $U_{\sigma},
t_{\sigma}, U_{am}$ and $\alpha$ from the microscopic Hamiltonian,
but rather treat them as free parameters. We take $U_a=U_m=1$ and $\alpha=0.5$,
and vary $U_{am}$ and $t_a=t_m$ since the qualitative aspects
of the phase diagram do not depend on the precise values of
$\alpha$, $U_a/U_m$ and $t_m/t_a$ \cite{note3}. We also neglect 
the trap potential, which is not expected to change qualitatively the
main conclusions of our work. 

Let us first discuss the phase diagram in the ``local''
(i.e. single-site) limit where $t_a=t_m=0$. For an integer number
$n_0$ of particles per site, the Hilbert space is
spanned by the states $|n_a,n_m\rangle = (n_a!n_m!)^{-1/2}
(\psi_a^\dagger)^{n_a} (\psi_m^\dagger)^{n_m}| {\rm vac}\rangle$ where
the number of atoms ($n_a$) and molecules ($n_m$) satisfy
$n_a+2n_m=n_0$. $|{\rm vac}\rangle$ denotes the vacuum of
particles. In this basis, the Hamiltonian $H^{(n_0)}$ is a 
real symmetric matrix, defined by  
\begin{eqnarray}
H^{(n_0)}_{n_m,n_m} &=& -\mu n_0 +\nu n_m+\frac{U_a}{2} n_a(n_a-1) 
+\frac{U_m}{2} n_m(n_m-1) + U_{am} n_a n_m ,  \nonumber \\ 
H^{(n_0)}_{n_m-1,n_m} &=& - \alpha [(n_a+2)(n_a+1)n_m]^{1/2} ,
\end{eqnarray}
which can be numerically diagonalized. In the following we shall
denote the eigenstates and eigenenergies by
$(\phi_i^{(n_0)}, E_i^{(n_0)})$, $i=1,\cdots, d_{n_0}$, where
$d_{n_0}=n_0/2+1$ ($(n_0+1)/2$) for $n_0$ even (odd). 

\begin{figure}
\epsfysize 3.7cm 
\epsffile[20 575 365 700]{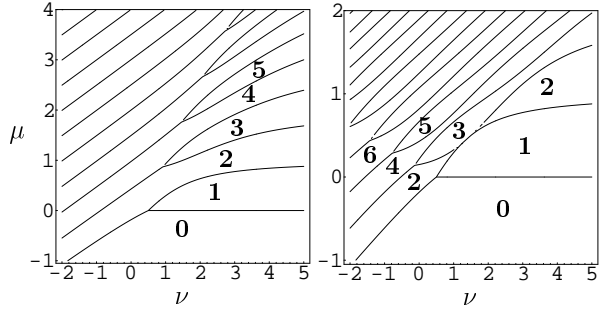}
\caption{Phase diagram in the local limit ($t_a=t_m=0$)  {\it vs}
  chemical potential $\mu$ 
  and detuning $\nu$ from the Feshbach resonance. $\alpha=0.5$,
  $U_a=U_m=1$. $U_{am}=1$ (left),  
  $U_{am}=0.25$ (right). Each phase is
  labeled by the number $n_0$ of atoms (free or bound into molecules)
  per site. } 
\label{fig1}
\end{figure}

The phase diagram in the local limit as a function of the
chemical potential $\mu$ and the detuning $\nu$ is shown in
Fig.~\ref{fig1} for $U_{am}=1$ and $U_{am}=0.25$. Each phase is
labeled by the total number $n_0=n_a+2n_m$ of bosons per site
where $n_a$ and $n_m$ are the mean number of atoms and molecules.
First let us consider the Mott states for large detuning
$|\nu|/\alpha \gg 1$. For even $n_0$, the Mott states are stable
on both sides of the resonance: for $\nu
> 0$, there are $n_a=n_0$ atoms per site whereas for $\nu < 0$
there are $n_m=n_0/2$ molecules per site. For $n_0$ odd, the MI's
are stable only for $\nu>0$. For $\nu<0$, the ground
state for odd $n_0$ is a superposition of two Mott phases with
$(n_0-1)/2$ and $(n_0+1)/2$ molecules per site and hence unstable
to superfluidity for any finite $t_a,t_m$. The phase diagram
for small detuning ($|\nu|/\alpha < 1$), on the other
hand, depends crucially on $U_{am}$. In contrast to the case
$U_{am} \ge 1$, for $U_{am} \ll 1$, the Mott phases with odd
$n_0$ can occur even when $\nu < 0$ as seen in Fig.~\ref{fig1}.

In order to study the instability of the Mott phases towards superfluidity
in the presence of finite intersite hopping, we perform an expansion
about the local limit. Our approach is a generalization of mean-field
theories previously used for the Bose-Hubbard model \cite{Sachdev99}.  
We write the partition function $Z$ as a functional integral over
complex bosonic 
fields $\psi_a,\psi_m$ with the action $S[\psi_a,\psi_m]=\int_0^\beta
d\tau \bigl\lbrace \sum_{{\bf r},\sigma} \psi^*_\sigma({\bf r})
\partial_\tau \psi_\sigma({\bf r}) + H[\psi] \bigr\rbrace$ ($\tau$ is
an imaginary time and $\beta=1/T$ the inverse temperature).
Introducing an auxiliary field $\phi_\sigma$  to decouple the intersite
hopping term by means of a Hubbard-Stratonovich transformation, we
obtain 
\begin{eqnarray}
Z &=& \int {\cal D}[\psi,\phi] e^{-(\phi|t^{-1}\phi) + 
((\phi|\psi) + {\rm c.c.}) -S_0[\psi] } \nonumber \\ 
&=& Z_0 \int {\cal D}[\phi]  e^{-(\phi|t^{-1}\phi)} 
\bigl\langle e^{(\phi|\psi) + {\rm c.c.}} \bigr\rangle_0
\nonumber \\ &=&  Z_0 \int {\cal D}[\phi]
e^{-(\phi|t^{-1}\phi) +W[\phi] } , 
\label{part}
\end{eqnarray}
where we use the shorthand notation $(\phi|\psi)=\int_0^\beta d\tau
\sum_{{\bf r},\sigma} \phi^*_\sigma({\bf r})\psi_\sigma({\bf r})$,
etc. $t^{-1}$ denotes the inverse of the intersite hopping matrix
defined by $t^\sigma_{{\bf r}{\bf r}'}=t_\sigma$ if ${\bf r},{\bf r}'$
are nearest neighbors and $t^\sigma_{{\bf r}{\bf r}'}=0$
otherwise. $S_0$ is the action of the $\psi$ field in the local
limit ($t_a=t_m=0$). $\langle \cdots \rangle_0$ means that the
average is taken with $S_0[\psi]$. In the last line of
(\ref{part}), we have introduced the generating functional of
connected local Green's functions \cite{Negele}:
\begin{eqnarray}
W[\phi] &=& -\sum_{\sigma,1,2} \phi^*_\sigma(1) G_\sigma(1,2)
\phi_\sigma(2) 
+ \frac{1}{4} \sum_{\sigma,1,2,3,4} G^{\rm IIc}_\sigma(1,2;3,4) 
\phi^*_\sigma(1)\phi^*_\sigma(2)\phi_\sigma(4)\phi_\sigma(3) 
\nonumber \\ && 
+  \sum_{1,2,3,4} G^{\rm IIc}_{am}(1,2;3,4)
\phi^*_a(1)\phi^*_m(2)\phi_m(4)\phi_a(3)  \nonumber \\ && 
+ \frac{1}{2} \sum_{1,2,3} [ \Lambda_{am}(1,2;3)
  \phi^*_a(1)\phi^*_a(2) \phi_m(3) + {\rm c.c.} ] ,
\label{W}
\end{eqnarray}
up to quartic order in the fields. Here $i\equiv ({\bf r}_i,\tau_i)$ and
$\sum_i \equiv \int_0^\beta d\tau_i \sum_{{\bf
  r}_i}$. $G_\sigma(1,2)=-\langle \psi_\sigma(1) \psi^*_\sigma(2)
\rangle_0$, $G^{\rm IIc}_\sigma(1,2;3,4) = \langle \psi_\sigma(1)
\psi_\sigma(2)\psi^*_\sigma(4)\psi^*_\sigma(3)\rangle_{0,c}$ and 
$G^{\rm IIc}_{am}(1,2;3,4) = \langle \psi_a(1)
\psi_m(2) \psi^*_m(4) \psi^*_a(3) \rangle_{0,c}$ are
single-particle and two-particle connected local Green's 
functions, and $\Lambda_{am}(1,2;3)=\langle \psi_a(1)\psi_a(2)
\psi^*_m(3)\rangle_0$. 

\begin{figure}
\epsfysize 3.2cm
\epsffile[-10 600 290 710]{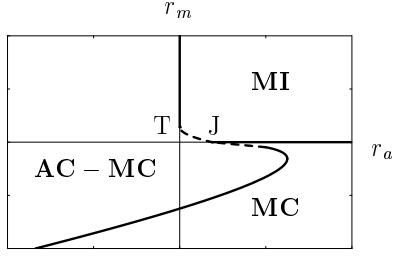}
\caption{Mean-field phase diagram obtained from the free energy $F$
  [Eq.~(\ref{free})]. MI: Mott insulator, AC-MC: SF phase with AC and MC, MC:
  SF phase with MC. The solid (dashed) lines indicate second (first) order
  transitions. $T=(0,\Lambda^2(g_m/g_a)^{1/2}/[4(g_ag_m)^{1/2}+
  g_{am}])$  and $J=(\Lambda^2/[4(g_ag_m)^{1/2}+ g_{am}],0)$ separate
  second and first order MI-SF transition lines.  } 
\label{fig2}
\end{figure}

We determine the zero-temperature phase diagram within a mean-field
approximation where $\phi_a$ and $\phi_m$ are assumed to be time and space
independent. A finite value of $\phi_\sigma$ signals superfluidity
since $\phi_\sigma=zt_\sigma\langle \psi_\sigma \rangle$ at the
mean-field level. The free energy (per site) $F$ is obtained to
quartic order from Eqs.~(\ref{part},\ref{W}): 
\begin{equation}
F = F_0 + \sum_\sigma \Bigl(r_\sigma \phi^2_\sigma +
  \frac{g_\sigma}{2} \phi^4_\sigma\Bigr) + g_{am} \phi^2_a\phi_m^2 -
  \Lambda \phi^2_a\phi_m,
\label{free}
\end{equation}
where, with no loss of generality, the order parameters can be chosen
real. $F_0=-T\ln Z_0$ is the free energy in the local
limit. The coefficients of the free energy expansion (\ref{free}) are 
given by $r_\sigma=(zt_\sigma)^{-1}+G_\sigma$, $g_\sigma=-G^{\rm
  IIc}_\sigma/2$, $g_{am}=-G^{\rm IIc}_{am}$, and
$\Lambda=\Lambda_{am}$ where all the local Green's functions are
evaluated in the static limit and for $T\to 0$. For instance $G_\sigma\equiv
G_\sigma(i\omega=0)$, where $G_\sigma(i\omega)$ is the Fourier transform of
$G_\sigma(\tau)$ ($\omega$ is a bosonic Matsubara frequency).
The mean-field phase diagram obtained from $F$ is shown in 
Fig.~\ref{fig2} \cite{Toledano}. There are three different ground
states: MI: $\phi_a=\phi_m=0$, SF phase with both an
atomic and a molecular condensate (AC-MC): 
$\phi_a,\phi_m\neq 0$, and SF phase with only a molecular condensate
(MC): $\phi_a=0,\phi_m\neq 0$. The existence of two different SF
phases in a Bose-Einstein condensate near a Feshbach resonance has
been recently predicted in 
Ref.~\cite{Rad04} \cite{note2}. Both SF phases spontaneously break
the U(1) gauge invariance, i.e. the invariance under the phase
transformations $\phi_a\to \phi_a e^{i\theta}$ and
$\phi_m \to \phi_m e^{2i\theta}$. The MC phase breaks
only the U(1)/Z$_2$ symmetry since $\phi_m
\to \phi_m $ for $\theta=\pi$. The
residual discrete Z$_2$ symmetry is spontaneously broken at the (Ising-like)
transition to the AC-MC phase which occurs near the resonance
\cite{Rad04}.
 
The phase transitions are second order except near $r_a=r_m=0$ where
they become first order (Fig.~\ref{fig2}) \cite{note5}. 
In the following, we neglect the possibility of first-order phase
transitions so that the phase boundaries of the Mott insulating phases
are determined by $r_a=0$ or $r_m=0$. This approximation
overestimates the stability of the Mott phases, but is expected to
give a correct qualitative description of the phase diagram. Since the
region where first-order phase transitions occur shrinks with $\Lambda$ (and
therefore $\alpha$), as shown in Fig.~\ref{fig2}, this approximation
should be quantitatively correct when $\alpha\ll U_a,U_m,U_{am}$. 

\begin{figure}
\epsfysize 3.7cm 
\epsffile[20 575 365 700]{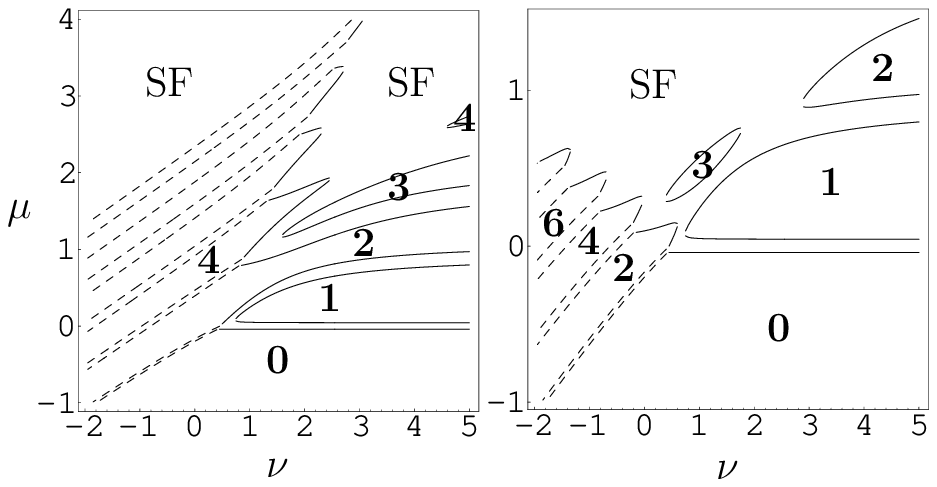}
\caption{Same as Fig.~\ref{fig1}, but for a finite intersite
 hopping $zt_a=zt_m=0.04$. The MI's (labeled by the number $n_0$ of atoms
 (free or bound into molecules) per site) are surrounded by SF phases. 
 The solid (dashed) lines indicate a
 transition from a MI to an AC-MC (MC) SF
 phase. Transition lines between AC-MC SF phases and MC SF phases are not
 shown. 
}
\label{fig3}
\end{figure}

To proceed further, we need to determine $r_\sigma$ and therefore the
single-particle local Green's function $G_\sigma$. Using the
eigenstates $\phi_i^{(n_0)}$ of $H^{(n_0)}$, we find (for $T=0$)
\begin{equation}
G_\sigma = - \sum_{i=1}^{d_{n_0-p_\sigma}}
\frac{|\langle \phi^{(n_0-p_\sigma)}_i |\psi_\sigma | \phi^{(n_0)}_1
  \rangle |^2}{E^{(n_0-p_\sigma)}_i-E^{(n_0)}_1} 
+  \sum_{i=1}^{d_{n_0+p_\sigma}}
\frac{|\langle \phi^{(n_0)}_1 |\psi_\sigma | \phi^{(n_0+p_\sigma)}_i
  \rangle |^2}{E^{(n_0)}_1-E^{(n_0+p_\sigma)}_i}  ,
\end{equation}
where $p_a=1$ and $p_m=2$, and $\phi^{(n_0)}_1$ denotes the ground
state of $H^{(n_0)}$. Fig.~\ref{fig3} shows the phase diagram in
the $(\nu,\mu)$ plane for $zt_a=zt_m=0.04$. Mott phases
with different $n_0$ are now separated by SF phases. Transitions between
AC-MC and MC phases, which are beyond the scope of this Letter, are
not shown. MI's with odd and even commensurate densities (to be
referred to as ``odd'' and ``even'' MI's in the following) exhibit
different behaviors. As $t_a=t_m$ increases, odd MI's 
become instable towards an AC-MC SF phase and are pushed to higher values of
$\nu$. Above the critical value $zt_a^{(c)}=U_a(2n_0+1+
\sqrt{n_0^2+n_0})^{-1}$, the odd MI with $n_0$ atoms per
site has become entirely superfluid. The behavior of even
MI's is more involved. They can be stable on both sides of the
Feshbach resonance, and are unstable both towards AC-MC and MC
phases as $\nu$, $\mu$ or $t_a=t_m$ are varied. The phase diagram and
the disappearance of the MI's as $t_a=t_m$ increases turns out to be
strongly dependent on the precise values of $U_{am}$ and $\alpha$, as
well as on $U_a/U_m$ and $t_a/t_m$. Quite generally, however, we find
that for a commensurate density and certain values of the strength of
the optical lattice the ground-state may undergo several
MI-SF or SF-MI transitions as $\nu$ is varied. This phenomenon always
occurs for even commensurate densities [see Fig.~\ref{fig3} for
$n_0=4$ (left panel) and $n_0=2$ (right panel)] but is restricted to
small values of $U_{am}$ for odd commensurate densities [right panel of
Fig.~\ref{fig3} for the $n_0=3$ phase]. 

\begin{figure}
\epsfysize 3.8cm 
\epsffile[25 580 365 698]{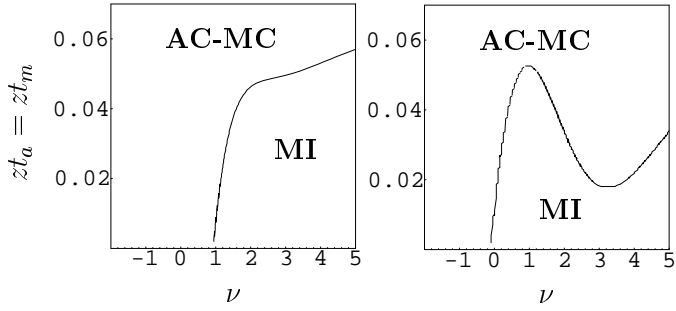}
\caption{Phase diagram for $n=3$ {\it vs} intersite hopping amplitude
  $t_a=t_m$ and detuning $\nu$ from the Feshbach resonance.
  $\alpha=0.5$ and $U_a=U_m=1$. $U_{am}=1$ (left), $U_{am}=0.25$ (right). }
\label{fig4}
\end{figure}

\begin{figure}
\epsfysize 3.8cm 
\epsffile[25 580 365 698]{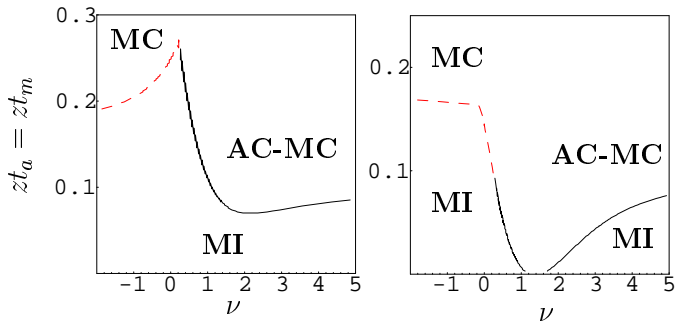}
\caption{Same as Fig.~\ref{fig4} but for $n=2$. The solid
  (dashed) lines indicate a transition from a MI to an AC-MC (MC) SF
   phase. } 
\label{fig5}
\end{figure}

The difference between odd and even phases is also clearly seen by
considering the phase diagram at constant density. The density
$n$ is given by $n=-\partial F/\partial\mu$. In the SF phase near the
SF-MI transition, we obtain
\begin{equation}
n = 
\left \lbrace 
\begin{array}{ll}  n_0 + \frac{r_a}{g_a-\frac{\Lambda^2}{2r_m}}
  \frac{\partial 
  r_a}{\partial \mu} & \,\,\,\, {\rm (AC-MC \,\, phase)} \\ 
n_0 +  \frac{r_m}{g_m} \frac{\partial r_m}{\partial \mu} &
\,\,\,\, {\rm (MC \,\, phase)} 
\end{array} \right. 
\end{equation} 
where $n_0$ is the density of the nearby MI. 
Thus at the transition points where $\partial_\mu r_\sigma=0$, the
density $n$ remains commensurate and equal to $n_0$. The condition
$\partial_\mu r_\sigma=0$ implies that the boundary between the MI and
the SF phases has a vertical tangent in the $(\nu,\mu)$ plane
(Fig.~\ref{fig3}). The phase diagrams for $n=3$ and $n=2$ are shown in
Figs.~\ref{fig4} and \ref{fig5}. As discussed above, we observe MI-SF
and SF-MI transitions as $\nu$ is varied at constant $t_a=t_m$. 
For large $U_{am}$, the MI will be more
robust for an integer mean number of atoms and molecules, whereas for small
$U_{am}$ it is more likely to become superfluid. This explains why
the behavior of the $n=2$ phase near $\nu\sim 1$ is strongly dependent
on $U_{am}$ (Fig.~\ref{fig5}).  

Let us finally discuss a few important points with respect to
experiments. In the vicinity of the Feshbach transition, where the interaction
energy becomes strong, the validity of a single-band Hubbard model becomes
questionable. However, this effect is significant only in a region of width
$\alpha/U$ around $\nu=0$ and will not change our conclusions for most regions
of the phase diagram for narrow resonances. 
Quantitative predictions of our model have to be supplemented with standard
T-matrix renormalization effects near the MI-SF transition. This requires a
more thorough analysis of the role of the optical lattice
\cite{Dickerscheid04} and is beyond the scope of our work. However, we expect
the qualitative predictions of the present theory regarding the Mott phases
and the MF-SI transitions to be valid. 
At high lattice filling, molecules are likely to be unstable due
to inelastic three-body collisions; this difficulty does not arise in the
$n=2$ and $n=1$ Mott phases, which are therefore good experimental candidates
to test our predictions. In the presence of the trap potential, superfluid
and Mott phases can coexist in certain ranges of experimental parameters, and
MI-SF transitions should rather be seen as crossovers than real quantum phase
transitions \cite{Wessel04}. However, this does not invalidate our main
conclusions as stated below (if we understand transitions as crossovers).  

In conclusion, we have shown that a Bose gas in an optical lattice near a
Feshbach resonance exhibits a very rich phase diagram. Characteristic
features of this phase diagram, which are qualitatively robust
irrespective of precise numerical parameter values, are: i) MI-SF transitions
can be induced 
either by decreasing the strength of the optical lattice or by
varying the magnetic field which controls the detuning from
resonance. ii) MI's with odd commensurate density are
always unstable to superfluidity on the molecular side of the
resonance whatever the strength of the optical lattice potential. iii)
MI's with even commensurate density can be stable on both sides of the
resonance. iv) For a commensurate density the ground-state may undergo 
several MI-SF or SF-MI transitions as the magnetic field is varied
through the resonance. 
We therefore expect a measurement of the momentum distribution function
\cite{Greiner02} to
display several transitions between peaked interference patterns (SF state)
and featureless uniform distribution (MI) as the magnetic field is
varied. Another possible experiment would be to measure the evolution
of the excitation spectrum as the magnetic field is varied, using
two-photon Bragg spectroscopy \cite{Kohl04}.

K.S. thanks S.M. Girvin for support during completion of this work.


\begin{thebibliography}{99}
\bibitem{note1} An important point in this derivation is that the size of
the molecular bound state is much smaller than the lattice period, so that the 
molecules can be considered as point-like particles of mass $2m$ (with
$m$ the atomic mass).
\bibitem{note3} Since the molecules are heavier than the atoms, we
  expect $t_m<t_a$.
\bibitem{note2} Note that there is no SF phase with an AC and no
  MC. Because of the coupling between atoms and molecules ($\alpha$
  term in Eq.~(\ref{free})), an AC ($\phi_a \neq 0$) would immediately induce a
  MC ($\phi_m\neq 0$).
\end{thebibliography}
\end{document}